\documentclass[a4paper]{jpconf}
\usepackage{graphicx}

\usepackage{amsmath}
\usepackage{amstext}
\usepackage{amssymb}
\usepackage{amsfonts}
\usepackage{bm} 
\usepackage{xfrac}
\usepackage{sidecap}
\usepackage{scalerel}

\begin{document}

\title{On the occurrence of buoyancy-induced oscillatory growth instability in directional solidification of alloys}

\author{Josep Maria Barbera$^{1,2}$, Thomas Isensee$^{1,3}$ and Damien Tourret$^1$}

\address{$^1$ IMDEA Materials, Getafe, Madrid, \sc Spain}
\address{$^2$ Universidad Politecnica de Madrid, ETSI Industriales, Madrid, \sc Spain}
\address{$^3$ Universidad Politecnica de Madrid, ETSI Caminos, Canales \& Puertos, Madrid, \sc Spain}

\ead{damien.tourret@imdea.org}

\begin{abstract}
Recent solidification experiments identified an oscillatory growth instability during directional solidification of Ni-based superalloy CMSX4 under a given range of cooling rates. 
From a modeling perspective, the quantitative simulation of dendritic growth under convective conditions remains challenging, due to the multiple length scales involved. 
Using the dendritic needle network (DNN) model, coupled with an efficient Navier-Stokes solver, we reproduced the buoyancy-induced growth oscillations observed in CMSX4 directional solidification. These previous results have shown that, for a given alloy and temperature gradient, oscillations occur in a narrow range of cooling rates (or pulling velocity, $V_p$) and that the selected primary dendrite arm spacing ($\Lambda$) plays a crucial role in the activation of the flow leading to oscillations. 
Here, we show that the oscillatory behavior may be generalized to other binary alloys within an appropriate range of $(V_p,\Lambda)$ by reproducing it for an Al-4at.\%Cu alloy.
We perform a mapping of oscillatory states as a function of $V_p$ and $\Lambda$, and identify the regions of occurrence of different behaviors (e.g., sustained or damped oscillations) and their effect on the oscillation characteristics.
Our results suggest a minimum of $V_p$ for the occurrence of oscillations and confirm the correlation between the oscillation type (namely: damped, sustained, or noisy) with the ratio of average fluid velocity $\overline V$ over $V_p$.
We describe the different observed growth regimes and highlight similarities and contrasts with our previous results for a CMSX4 alloy.
\end{abstract}

\section{Introduction}

Directional solidification (DS) is critical to manufacture a range of high-value components. 
A prominent example is that of single-crystal Ni-based superalloy turbine blades for aeronautical applications \cite{pollock2006nickel}. 
They are arguably among the most advanced manmade structural metallic components, capable to withstand extreme thermal, mechanical, and chemical conditions all at once.
However, various defects may emerge, such as segregated channels and freckles \cite{copley1970origin,sample1984mechanisms,hellawell1993channel,pollock1996breakdown}, which may lead to a significant amount of manufactured parts being discarded.
Therefore, a better fundamental understanding of the conditions leading to stable and homogeneous DS is critical to the production of the next-generation of directionally solidified components.

Recently, experiments using X-ray {\it in situ} radiography revealed the existence of oscillatory instabilities during DS of Ni-based superalloy CMSX4 \cite{reinhart2020impact}. 
Over the years, a range of oscillatory instabilities during DS had been previously investigated, e.g., related to growth in a narrow channel \cite{karma1989oscillatory}, coordinated ``breathing modes'' in cellular array growth \cite{georgelin1997oscillatory,bergeon2013spatiotemporal,tourret2015oscillatory}, sidebranching at high primary spacing \cite{echebarria2010onset}, or banding instabilities in rapid solidification \cite{kurz1996banded,ji2022microstructural}. 
However, while all of these naturally occur within a purely diffusive thermo-solutal transport regime, the oscillations observed in CMSX4 stem from gravity-induced buoyancy \cite{reinhart2020impact}, making them more challenging to investigate.
The effect of buoyant convection on dendritic growth and microstructure heterogeneities has been acknowledged for decades \cite{mehrabian1970interdendritic, dupouy1989natural, jamgotchian2001localized, bogno2011analysis, shevchenko2013chimney}.
However, due to the various scales to be incorporated together in the problem, quantitative modeling of dendritic growth in presence of fluid flow remains computationally demanding \cite{sakane2017multi, sakane2020two}.

Using a multiscale dendritic needle network (DNN) approach \cite{tourret2013multiscale,tourret2016three}, extended to include fluid flow in the liquid phase \cite{tourret2019multiscale,isensee2022convective}, we recently managed to reproduce the growth oscillations observed in CMSX4 \cite{isensee2022convective}. 
Beyond reproducing experimental observations --- e.g., the transition from oscillatory to damped oscillations when increasing the cooling rate (or equivalent pulling velocity $V_p$) --- our simulations highlighted the importance of the primary dendritic spacing $\Lambda$ on the occurrence (at high $\Lambda$) or inhibition (at low $\Lambda$) of these oscillations, and the fact that sustained oscillations seemed to occur when the average fluid velocity $\overline V$ was close to $V_p$.

Computational simulations of this oscillatory behavior naturally open the way for a deeper and systematic exploration of the conditions of occurrence and characteristics of oscillations as a function of alloy parameters and processing conditions. 
Extending our previous study \cite{isensee2022convective}, here we perform a mapping of this buoyancy-induced oscillatory behavior as a function of primary spacing and pulling velocity. 
An underlying objective is also to uncover whether this phenomenon is generalizable to other alloys, provided the appropriate combination of temperature gradient $G$, pulling velocity $V_p$, and primary spacing $\Lambda$. 
Hence, we focus on a different alloy with relatively well-known phase diagram and thermophysical properties, namely Al-4at.\%Cu.

\section{Methods}

The model, its implementation, and the simulations are similar in nature to those presented elsewhere \cite{isensee2022convective,tourret2016three,tourret2019multiscale}. Therefore, here we only summarize their main features (see Ref.~\cite{isensee2022convective} for further details).

\subsection{Model}

The dendritic needle network (DNN) model alleviates the burden of explicitly tracking the morphologically complex solid-liquid interface by representing growing crystals as hierarchical networks of thin parabolic-shaped branches \cite{tourret2016three,tourret2019multiscale,isensee2022convective}. 
The temperature field, $T(\mathbf x,t)$, is usually imposed as a boundary condition throughout the domain, and the evolution of the solute concentration field, $c(\mathbf x,t)$, is solved considering either diffusive \cite{tourret2016three} or convective \cite{tourret2019multiscale,isensee2022convective} conditions in the liquid phase, while the concentration along the solid-liquid interface, i.e., along the needle network, is set at the equilibrium concentration at the local temperature.

We consider a dilute binary alloy with constant liquidus slope, $m<0$, and solute partition coefficient, $0<k<1$, and introduce the reduced (dimensionless) concentration field, $U \equiv (c^0_l-c)/[(1-k)c^0_l]$, where $c_l^0$ is the liquid equilibrium concentration at the reference (i.e., solidus) temperature, $T_0$. 
Directional solidification conditions are represented via the frozen temperature approximation $T=T_0+G(x-V_p t)$, where $G$ is the strength of the temperature gradient and $V_p$ is the pulling (or isotherm) velocity, both imposed along the $x$ direction. 
Neglecting curvature and kinetic undercooling contributions, the equilibrium dimensionless concentration, $U_i$, along the solid-liquid interface is expressed as 
\begin{equation}
\label{eq:equilU}
 U_i = (x-V_p t)/l_T,
\end{equation}
with $l_T = |m|(1-k)c^0_l/G$ the thermal length separating liquidus and solidus temperatures.

The liquid velocity field, $\bm{v}$, is calculated using Navier-Stokes equations  
\begin{equation}
\label{eq:ns}
\rho\left[\partial_t\bm{v}+(\bm{v}\cdot\nabla)\bm{v}\right] = \bm{F} - \nabla p + \eta\nabla^2\bm{v} , 
\end{equation}
where $\rho$ is the fluid density, $p$ its pressure, $\eta$ its viscosity and $\bm{F}$ corresponds to external forces. 
The liquid is assumed incompressible with $\nabla\cdot\bm{v}=0$ and a null velocity is imposed along the solid-liquid interface, i.e., along the needles.
Buoyant forces are included using the Boussinesq approximation, considering only solute-dependent terms 
\begin{equation}
 \bm{F} = \rho^l_\infty\bm{g}\left[1-\beta_c(c-c_\infty)\right],
\end{equation}
where  $\beta_c \equiv -(\partial\rho/\partial c|_{c=c_\infty})/\rho^l_\infty$ is a solutal expansion coefficient, with $\rho^l_\infty$ the fluid density at the alloy nominal concentration $c_\infty$ and $\bm{g}$ is the gravity acceleration.
The transport of solute in the liquid with fluid velocity $\bm{v}$ is thus described by the advection-diffusion equation, with diffusion coefficient $D$,
\begin{equation}
\label{eq:U}
 \partial_t U + \nabla\cdot(\bm{v}U) = D\nabla^2 U.
\end{equation}

The instantaneous tip radius, $R(t)$, and velocity, $V(t)$, of each needle-like branch is computed using the microscopic solvability condition
\begin{equation}
\label{eq:r2v}
 R^2 V = 2Dd_0 \big/ \big\{[1-(1-k)U_t]\sigma\big\},
\end{equation}
where $d_0 = \Gamma/\left[|m|(1-k)c_l^0\right]$ is the capillary length at $T_0$ with $\Gamma$ the interface Gibbs-Thomson coefficient, $U_t = (x_t-V_p t)/l_T$ is the equilibrium concentration at the tip position, $x_t$, and $\sigma$ is the tip selection parameter \cite{langer1980instabilities,kurz2019progress}, combined with a solute conservation statement in the vicinity of the parabolic tip
\begin{equation}
\label{eq:rv2}
 RV^2 = 2D^2\mathcal{F}^2 \big/ \big\{ \left[1-(1-k)U_t\right]^2 d_0 \big\},
\end{equation}
where the flux intensity factor $\mathcal{F} \equiv \int_{\Gamma_0} (\partial_n U)\,dS/\big(4\sqrt{a/d_0}\big)$ measures the incoming flux and can be calculated along any contour $\Gamma_i$ (instead of directly on the solid-liquid interface along $\Gamma_0$) as 
\begin{equation}
\label{eq:fif}
  4\mathcal{F}\sqrt{a/d_0} = \scaleobj{.8}{\int_{\Gamma_i}} \left(\partial_{n} U\right) {\rm d}S +\frac{V}{D}\scaleobj{.8}{\int_{\Sigma_i}}\left(\partial_x U\right) {\rm d}A,
\end{equation}
with $\partial_{n} U$ the outward normal solute gradient, $\Sigma_i$ the surface enclosed between $\Gamma_0$ and $\Gamma_i$, and $a$ the distance between the tip and the intersection of the integration contour with the parabolic tip (i.e., the location where $\Gamma_0$ and $\Gamma_i$ meet) \cite{tourret2016three,tourret2019multiscale}.

The model is solved similarly as in Refs~\cite{tourret2019multiscale,isensee2022convective}, i.e., using a mostly explicit finite difference scheme on a staggered grid, an upwind discretization scheme for convective terms, a projection method for the resolution of the Navier-Stokes equations \cite{chorin1968numerical}, and an iterative successive over-relaxation (SOR) method \cite{frankel1950convergence, young1954iterative} for the incompressibility condition.
The code is implemented in C-based CUDA language to leverage acceleration using Graphics Processing Units (GPUs).

\subsection{Simulations}

In Ref.~\cite{isensee2022convective}, our simulations aimed at approaching specific experimental conditions of CMSX4 directional solidification \cite{reinhart2020impact}.
In contrast, here, an underlying objective is to investigate whether the oscillatory behavior may be general to any arbitrary alloy, provided the appropriate set of $(G,V_p,\Lambda)$ conditions.
Therefore, we consider a different alloy, namely a binary Al-4at.\%Cu, with relatively well known parameters \cite{isensee2022convective,steinbach2009pattern}, but so far no report of such oscillatory growth behavior. 
Specifically, we use alloy parameters as in Refs~\cite{isensee2022convective} (Section~3.1 therein) and \cite{steinbach2009pattern}, i.e., $c_\infty=4.0\,$at.\%, $D=3\times10^{-9}~$m$^2$/s, $k=0.14$, $m=1.6~$K/at.\%, $\nu=\eta/\rho=5.7\times10^{-7}~$m$^2$/s, $\Gamma=2.4\times10^{-7}~$Km, $\beta_c=-10^{-2}$/at\%, and $\sigma\approx0.153$ (corresponding to an interfacial excess free energy anisotropy $\epsilon=0.02$).
In an Al-Cu alloy (unlike in CMSX4), buoyant currents and plumes are caused by the rejection of the heavier element (Cu).
Therefore, while the growth direction is kept as $x+$ and represented upwards for consistency with Ref.~\cite{isensee2022convective}, gravity forces also have a $x+$ direction (i.e. pointing upwards in figures below).
We fix the temperature gradient at $G=10~$K/mm, and explore velocities from $V_p=40$ to 100~\textmu m/s within a range of primary dendrite arm spacing $\Lambda$ from about 90 to 350~\textmu m.

The radius of integration of the flux intensity factor is set to $r_i/R_s=4$ and the truncation radius of needles far behind the tip to $r_{\rm max}/R_s=5$, with $R_s$ the theoretical steady state tip radius in the diffusive regime \cite{tourret2016three,isensee2022convective}.
For each velocity $V_p$, the grid spacing is chosen between $\Delta x/R_s=0.7$ and 1.24, ensuring that the diffusion length included a sufficient number of grid points --- namely with $D/V_p/\Delta x$ ranging from 6 ($V_p=100$~\textmu m/s) to 16 ($V_p=40$~\textmu m/s).
Other numerical parameters are similar to those listed in Ref.~\cite{isensee2022convective} (Table~1 therein, with $K_{\Delta t}=0.5$).

With these parameters, the simulations are similar in nature to those presented and discussed in Ref.~\cite{isensee2022convective} (Sections~3.1 and 4 therein).
Within a domain of height $H$ ($x$-direction) and width $W$ ($y$-direction), we set an array of $N$ evenly spaced primary needles at the bottom of the domain growing in the $x+$ direction and with their tips initially aligned in $x$ along the liquidus temperature location.
Boundary conditions (BCs) are periodic laterally (in the $y$ direction) for all fields. No-flux ($\partial U/\partial x=0$) and free-slip ($v_x=0$) conditions are applied along the top and bottom boundaries (in the $x$ direction).
The simulations use a moving frame in the $x$-direction, such that the most advanced needle tip in $x$ remains at a fixed location --- namely at a distance between 30\% and 65\% of the domain height from the bottom boundary.

The domain size in $x$, i.e., its height $H$, is chosen long enough for BCs to have negligible effect on the flow pattern --- typically adjusted by trial-and-error, ensuring that the liquid length is always greater than $5D/V_p$ and the solid length greater than $8D/V_p$.
The domain size in $y$ is set to either 630 or 1470 inner grid points, for convenience (both numbers are divisible by a broad range of integer needle number $N$, as required to impose periodic arrays) and performance (accounting for the two extra points used to impose periodic BCs, both 632 and 1472 are divisible by 8, thus allowing nearly optimal GPU block sizes).
Each simulation is initialized with between 7 and 30 needle-like primary dendrites, in order to probe different values of primary spacing $\Lambda$.
The array growth was simulated for a duration of between 90 (high $V_p$) and 120 (low $V_p$) seconds, so as to obtain enough oscillation periods to analyze, when relevant.
Main parameters of the resulting 44 simulations are summarized in Table~\ref{tab:map}.
Each simulation is performed using a single GPU (Nvidia RTX3090, in most cases) and each required at most 92~h to complete (wall time).

\begin{table}[!h]
\caption{\label{tab:map} Simulation parameters for the mapping of $V_p$ and $\Lambda$.}%
\begin{center}
\begin{tabular}{ccccccc}
\\[-0.4in]
\br
Velocity & \multicolumn{2}{c}{Grid spacing}  & Height  & \multicolumn{2}{c}{Number of needles, $N$} & PDAS range\\
$V_p$ (\textmu m) & $\sfrac{\Delta x}{R_s}$ & $\Delta x$ (\textmu m) & $\sfrac{H}{\Delta x}$ & $\sfrac{W}{\Delta x}=630$ & $\sfrac{W}{\Delta x}=1470$ & $\Lambda$ (\textmu m) \\
\mr
40 & 1.24 & 7.06 & 638 & 14, 18, 21, 30 & - & (148, 318) \\
50 & 1.10 & 5.56 & 638 & 10, 14, 15, 18, 21, 30 & - & (117, 350) \\
60 & 1.00 & 4.59 & 510 & 9, 10, 14, 15, 18, 21, 30 & - & (96, 321) \\
70 & 0.94 & 3.98 & 510 & 7, 9, 10, 14, 15, 18, 21 & - & (119, 358) \\
80 & 0.90 & 3.54 & 398, 510 & 7, 9, 10, 14, 15, 18, 21 & - & (106, 319) \\
90 & 0.80 & 2.96 & 398, 510 & 7, 9, 10, 14, 15, 18 & 14 & (103, 310) \\
100 & 0.70 & 2.44 & 398 & 7, 9, 10, 14, 18 & 14 & (86, 257) \\
\br
\end{tabular}
\end{center}
\end{table}

\subsection{Post-processing}
\label{sec:postproc}

In order to classify the simulated growth behaviors and extract oscillation characteristics, we analyzed the velocity evolution, $V(t)$, of every needle tip in each simulation.
First, we discarded simulations leading to the elimination of one or more dendrites, which is expected to happen when $\Lambda$ is too low due to intense solutal interaction among dendrites \cite{echebarria2010onset,tourret2013multiscale,bellon2021multiscale}.
All remaining simulations thus retain a homogeneous spacing $\Lambda$ above the lower stability limit $\Lambda_{\rm min}$.

\begin{SCfigure}[][!h]
  \caption{
Illustration of functions $v_1(t)$  and $v_2(t)$ for $V_0=8.0$, $A=1.5$, $f=2.0$, $t_0=0.0$ and different values of $\tau$ (Eq.~\eqref{eq:decay}) or $S$ (Eq.~\eqref{eq:spiky}).\\~\\
  \label{fig:functions}
  }
\includegraphics[width=.575\textwidth, trim=0 0 -15 0,clip]{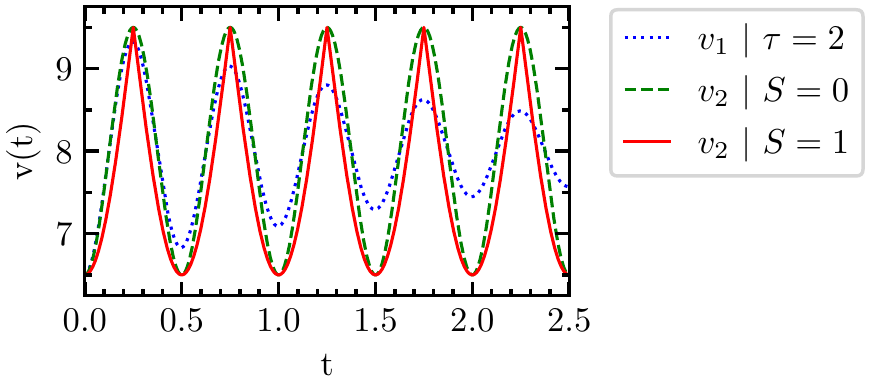}%
\end{SCfigure}

Individual needle velocities, $V(t)$, were then fitted to either of these two functions
\begin{align}
v_1(t) &= V_0 - A \cos\big[2\pi(t - t_0)f\big] \exp(-t/\tau) ,
 \label{eq:decay} \\
v_2(t) &=  V_0 - A\big\{ (1-S)\cos\big[2\pi(t-t_0)f\big]+2S\big| \cos\big[\pi(t-t_0)f\big] \big| - S \big\} ,
 \label{eq:spiky}
\end{align}
via the following fitting parameters: mid-range velocity $V_0$, time origin $t_0$, oscillation amplitude $A$, oscillation frequency $f$, and either a characteristic damping time $\tau$ (Eq.~\eqref{eq:decay}) or a ``spikiness'' factor $S$ (Eq.~\eqref{eq:spiky}).
The latter, bounded to $0\leq S\leq1$, allows fitting signals showing burst-like ``spiky'' maxima, yet with stable-frequency oscillations.
As illustrated in Figure~\ref{fig:functions}, $S=0$ leads to a regular cosine function and $S=1$ to a spiky cosine-like function of similar amplitude and frequency. 
(Note that, when $S>0$, the mid-range velocity $V_0$ differs from the average velocity $V_p$.)
We chose to fit to both functions and select the best fit out of the two for convenience, after noticing that damped oscillations typically had smooth (non-spiky) behaviors while sustained oscillations had a range of behaviors from smooth to spiky.
For each needle, the fitting was performed using a custom Python script that dynamically scans different fitting ranges and selects the one leading to the best fit, i.e., the highest coefficient of determination $r^2$, while ensuring that the range included a minimum of 5 oscillation periods (when relevant) or at least 20~seconds.
Since raw $V(t)$ data may be prone to numerical oscillations as the needle progresses through the grid \cite{tourret2019multiscale}, in order to facilitate the fitting, the velocities were smoothed applying a moving average to the time derivative of the needle length prior to fitting (both raw and smooth $V(t)$ are shown in later figures).
The initial guess for the frequency $f$, prior to its fitting, was based on a preliminary estimate via fast Fourier transform of the $V(t)$ signal.

\sidecaptionvpos{figure}{c}

Growth velocities $V(t)$ leading to a poor fit to both functions, namely if $r^2<0.8$, where classified as {\it noisy} (rather than rigorously {\it non-oscillatory} since their behavior was usually closer to noisy oscillations than to erratic bursts as reported in Ref.~\cite{isensee2022convective}).
When $r^2\geq0.8$, behaviors were classified as {\it damped} if the fit was better with Eq.~\eqref{eq:decay} than Eq.~\eqref{eq:spiky}, or {\it sustained} otherwise.
(Damped oscillations were all quite straightforward to identify from visual inspection of $V(t)$.)

\section{Results and Discussion}

\begin{SCfigure}[][!h]
  \caption{
Oscillatory behavior map.
Symbols types denote different growth behaviors: ($\times$) elimination events for $\Lambda<\Lambda_{\rm min}$ (black line) or ($\blacktriangledown$) damped, ($\bullet$) sustained, or ($\blacktriangle$) noisy oscillations.
Symbols are colored according to the ratio $\overline V/V_p$. 
Gray text labels next to symbols mark cases highlighted in the following figures and discussion.
\\~\\
  \label{fig:map}
  }
\includegraphics[width=.66\textwidth, trim= 0 0 -35 0, clip]{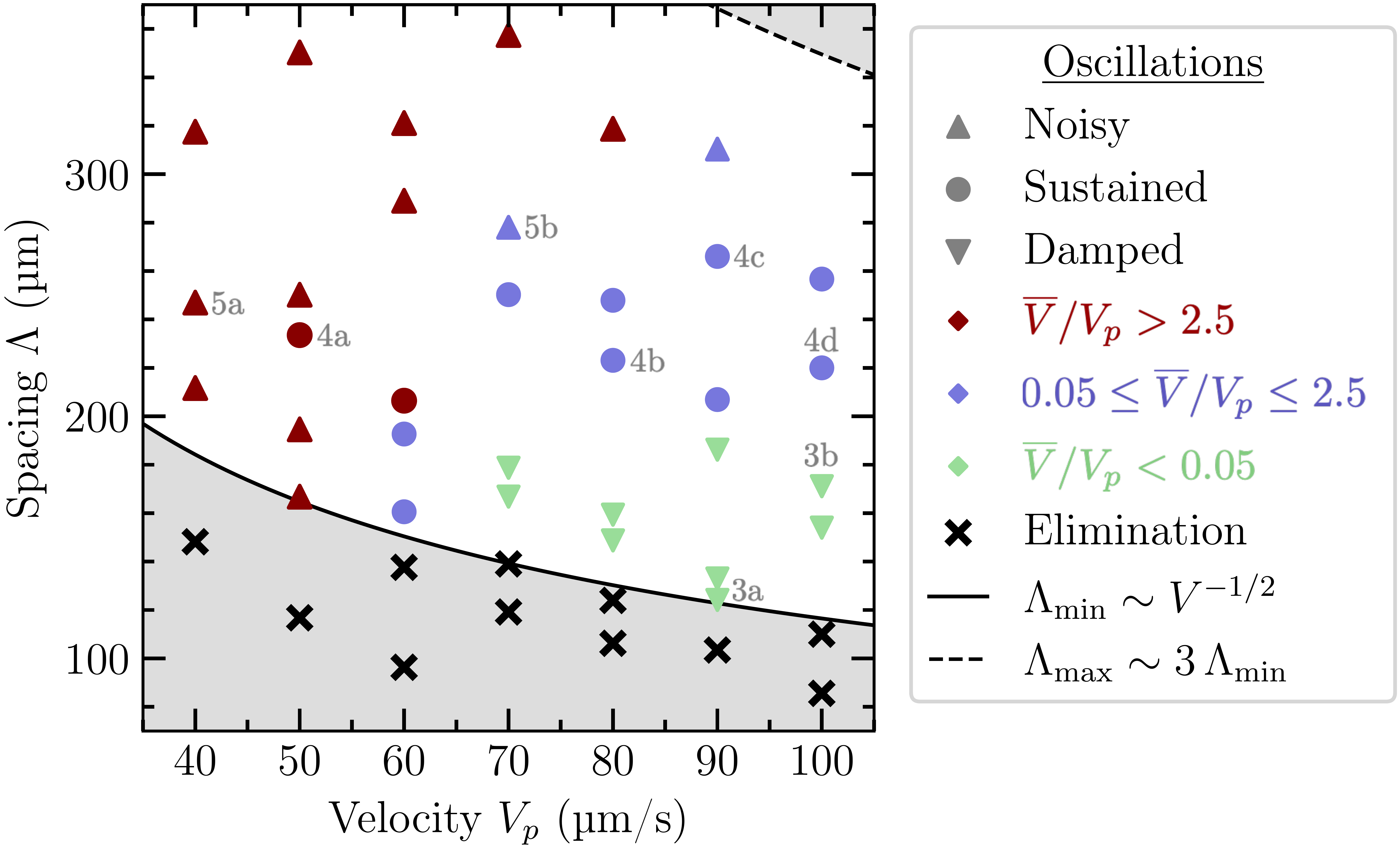}%
\end{SCfigure}

Figure~\ref{fig:map} shows the resulting map of growth behaviors as a function of pulling velocity $V_p$ and primary spacing $\Lambda$.
As a first conclusion, it was indeed possible to obtain sustained oscillatory growth for several simulations over a range of $(V_p,\Lambda)$. 
Oscillations were only found above a minimum velocity, here for $V_p\ge60~$\textmu m/s.
Most cases (all but one) leading to sustained oscillations ($\bullet$ symbols) fall within a spacing range $190\leq\Lambda/$\textmu m $\leq270$, while damped oscillations ($\blacktriangledown$ symbols) emerge within $120\leq\Lambda/$\textmu m $\leq190$.
While different symbol types denote the classification criteria mentioned in Section~\ref{sec:postproc}, their color follows a discrete distribution as a function of the ratio of average fluid velocity $\overline V$ over pulling velocity $V_p$, namely: light green ($\overline V/V_p\leq0.05$), medium blue ($0.05\leq\overline V/V_p\leq2.5$), or dark red ($\overline V/V_p\geq2.5$). 
Except for a few data points around the sustained/noisy transition, the reasonable match between symbol types and colors shows that we can correlate the occurrence of damped, sustained, or noisy oscillations to the ratio $\overline V/V_p$.
However, the current threshold values for $\overline V/V_p$, here denoted $\xi_1\approx0.05$ (damped/sustained) and $\xi_2\approx2.5$ (sustained/noisy),  differ from those identified in our previous study focused on CMSX4 ($\xi_1\approx 1.0$, $\xi_2\approx 2.0$), such that they may depend upon alloy parameters and/or processing conditions (e.g., temperature gradient).
Below, we describe and discuss in further details the different growth behaviors observed in our simulations.

\sidecaptionvpos{figure}{c}

\underline{Elimination ($\Lambda<\Lambda_{\rm min}$)}.
As expected, for each $V_p$, the lowest spacings lead to the elimination of at least one primary dendrite, marking the lower limit of the stable spacing range, $\Lambda_{\rm min}$.
This minimal stable spacing appears to reasonably match the expected power law $\Lambda_{\min}\sim V^{-1/2}$ (solid black line).
The ratio between the lower and upper limits of the stable spacing range $\Lambda_{\max}/\Lambda_{\min}$ typically varies between 2 and 5 \cite{echebarria2010onset,tourret2013multiscale,bellon2021multiscale}.
Since the current simulations do not include sidebranching, they cannot predict the upper limit (as done, e.g., in Refs~\cite{tourret2013multiscale,bellon2021multiscale}).
Hence, once we had an estimate of $\Lambda_{\min}$ for a given $V_p$, we approximated $\Lambda_{\max}\approx 3\,\Lambda_{\rm min}$ (black dashed line) and did not perform any simulation above this $\Lambda_{\max}$, because such a situation would in reality likely lead to spacing reduction by tertiary sidebranching \cite{tourret2013multiscale,bellon2021multiscale}.

\begin{figure}[!h]%
\centering{\includegraphics[width=.935\textwidth]{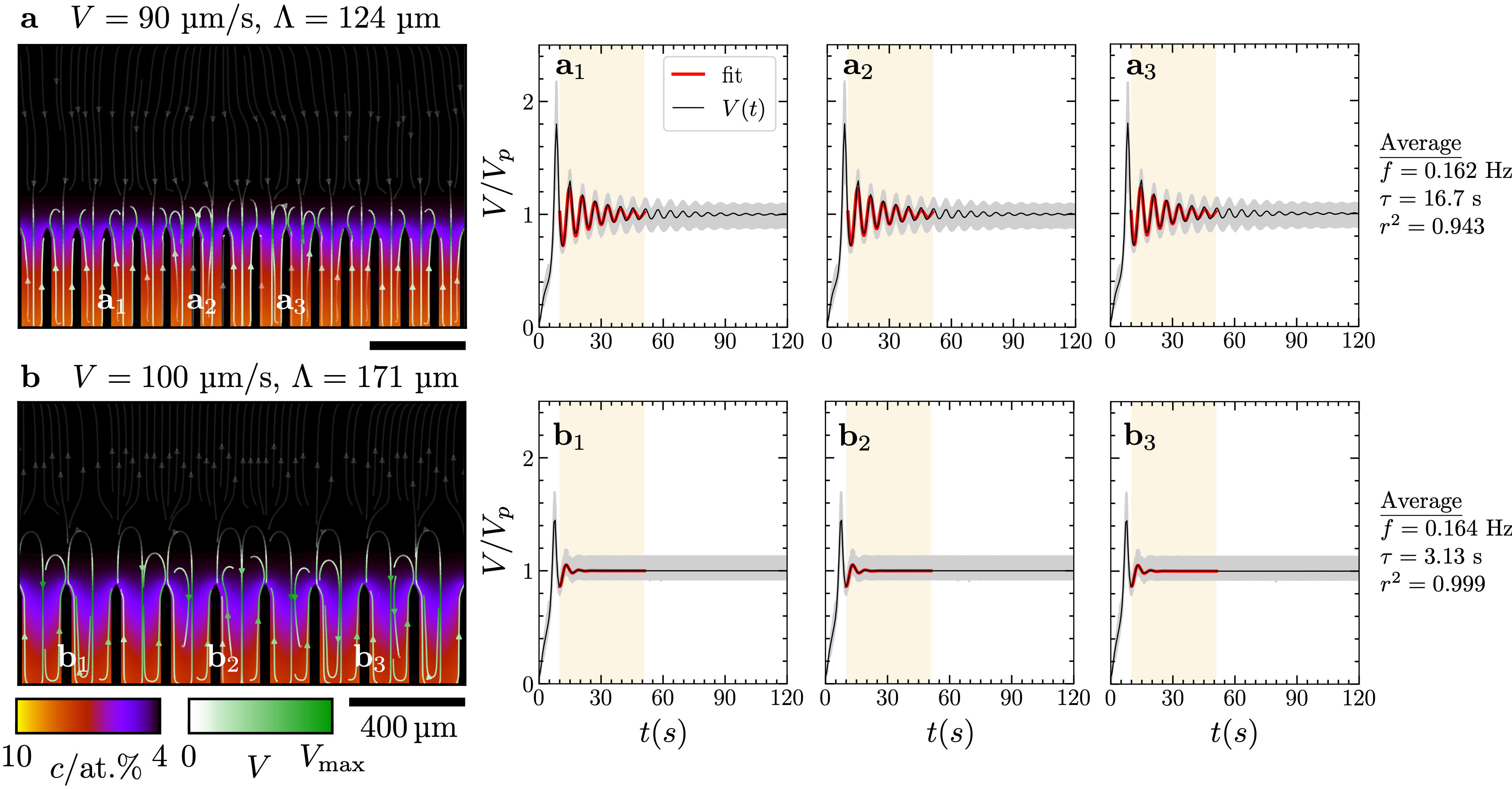}}%
\caption{
\underline{Damped oscillations} (a) for $V=90~$\textmu m/s and $\Lambda=124~$\textmu m and (b) for $V=100~$\textmu m/s and $\Lambda=171~$\textmu m.
(left) Concentration field (color map) and fluid flow (streamlines) at $t=120~$s.
Streamlines are progressively transparent toward lowest velocities (from opaque at $V/V_{\rm max}\geq0.2$ to 90\% transparent at $V=0$).
(right) Tip velocities of selected needles, showing raw (gray) and smoothed (black) $V(t)$ as well as fitted function (thick red) and range (shaded background).
  \label{fig:damped}
}%
\end{figure}

\underline{Damped oscillations ($\Lambda\gtrsim\Lambda_{\rm min}$)}. 
Above a certain velocity, here $\approx75~$\textmu m/s, the lowest stable spacings just above $\Lambda_{\rm min}$ lead to damped oscillations.
These are illustrated in Figure~\ref{fig:damped} for two representative case, namely for (a) $V=90~$\textmu m/s and $\Lambda=124~$\textmu m/s and for (b) $V=100~$\textmu m/s and $\Lambda=171~$\textmu m.
They respectively correspond to the lowest ($\tau=16.7~$s) and highest ($\tau=3.13~$s) damping rates obtained among the 44 simulations.
While we did not collect sufficient data to extract meaningful scaling laws, we observed that the damping rate typically increases (i.e., $\tau$ decreases) when either $V_p$ or $\Lambda$ increases.
In this regime, the behavior of all needles in the array is homogeneous and synchronized (typically oscillating in phase with one another).
From Figure~\ref{fig:damped} on, velocity fields are shown via the flow streamlines, of which the opacity decreases progressively when the fluid velocity is lower than $0.2V_{\rm max}$, with $V_{\rm max}$ the maximum fluid velocity over the entire domain. 
This allows illustrating that, within the damped oscillation regime (Fig.~\ref{fig:damped}), while some convective currents appear around and between the dendrites, the fluid velocity vanishes within a narrow boundary layer ahead of the solidification front. 
Transport of solute beyond this boundary layer occurs then primarily through diffusion.

\begin{figure}[!b]%
\centering{\includegraphics[width=.935\textwidth]{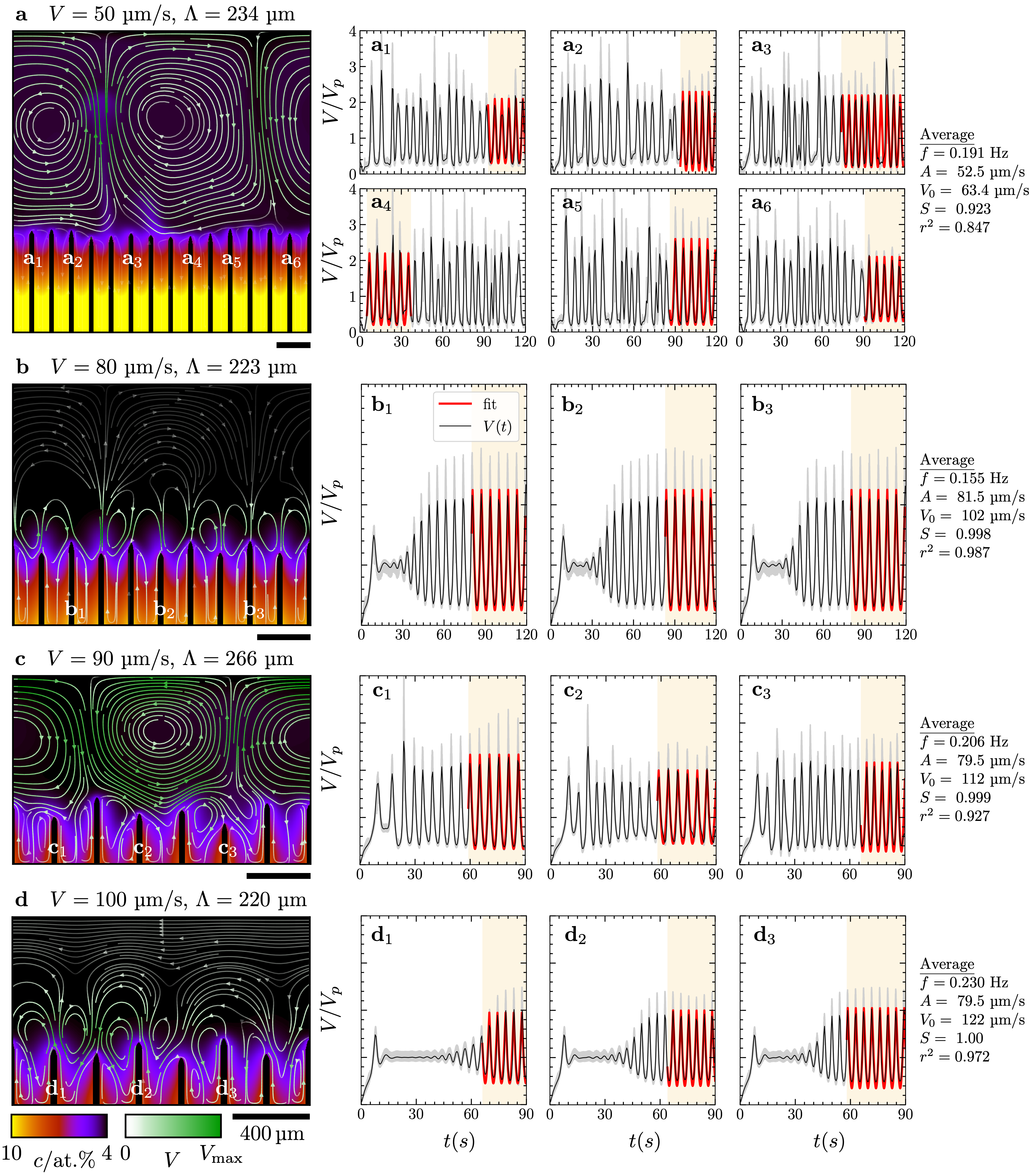}}%
\caption{
\underline{Sustained oscillations} for (a) $V=50~$\textmu m/s and $\Lambda=234~$\textmu m, (b) $V=80~$\textmu m/s and $\Lambda=223~$\textmu m, (c) $V=90~$\textmu m/s and $\Lambda=266~$\textmu m, (d) $V=100~$\textmu m/s and $\Lambda=220~$\textmu m.
(left) Concentration field (color map) and fluid flow (streamlines, from opaque at $V/V_{\rm max}\geq0.2$ to 90\% transparent at $V=0$) at $t=$ (a,b) 120~s and (c,d) 90~s.
(right) Tip velocities of selected needles, showing raw (gray) and smoothed (black) $V(t)$ as well as fitted function (thick red) and range (shaded background).
  \label{fig:sustained}
}%
\end{figure}

\underline{Sustained oscillations (intermediate $\Lambda$)}. Within the inspected range, for a sufficient pulling velocity, at intermediate primary spacings $\Lambda$, sustained oscillations occur. Typical cases are illustrated in Figure~\ref{fig:sustained} for (a) $V=50~$\textmu m/s and $\Lambda=234~$\textmu m, (b) $V=80~$\textmu m/s and $\Lambda=223~$\textmu m, (c) $V=90~$\textmu m/s and $\Lambda=266~$\textmu m, and (d) $V=100~$\textmu m/s and $\Lambda=220~$\textmu m.
Among these, Fig.~\ref{fig:sustained}c is a good illustration of a typical sustained oscillatory regime. 
Therein, all needles exhibit an oscillatory growth, with a small variability of oscillation amplitudes and frequencies throughout the array. 
As seen on the leftmost panel, prominent convection vortices appear in the liquid, and needle oscillations are desynchronized (not in phase) with one another.

In some cases, illustrated by Fig.~\ref{fig:sustained}b and d, the oscillatory growth regime emerges after a usually short (b) but potentially long (d) transient period, momentarily more akin to a damped oscillation regime before a subsequent amplification of the oscillation amplitude toward a steady value.
(For this reason, several runs classified as damped were performed for longer durations in order to ascertain that they were indeed not in a transient state.)
Such simulations exhibiting a transient regime were observed for the lowest $\Lambda$ data point classified as sustained at V = 80, 90, and 100~\textmu m/s, i.e., close to the edge of the transition between damped and sustained oscillations (Fig.~\ref{fig:map}).
The corresponding convective patterns are also intermediate, with vanishing velocities toward the top of the domain (like for damped oscillations at lower $\Lambda$) and the emergence of small nascent convection rolls (like for sustained oscillations at higher $\Lambda$).

Finally, Fig.~\ref{fig:sustained}a shows the other end of the sustained oscillation spectrum, at the edge of the transition toward noisy oscillations.
This specific case corresponds to the lone outlier data point classified as ``sustained'' for $V=50~$\textmu m/s in Fig.~\ref{fig:map}, due to its $r^2=0.847$ higher than the chosen threshold of 0.8.
However, in spite of a relatively good fit to $v_2(t)$, Fig.~\ref{fig:sustained}a shows that resulting $V(t)$ look relatively noisy (see, e.g., Fig.~\ref{fig:sustained}a$_3$).
In such cases, strong convection vortices are present within the liquid.

Among the sustained oscillations, measured frequencies range from 0.155 to 0.230~Hz (i.e., periods from 4.35 to 6.45~s), without any conclusive dependence upon $\Lambda$ or $V_p$ emerging within the investigated range of conditions.
Naturally, the oscillation amplitude scales approximately like $V_p$ and tends to increase toward high $\Lambda$ when the $V(t)$ signals become more spiky.

\clearpage

\begin{figure}[!h]%
\centering{\includegraphics[width=.935\textwidth]{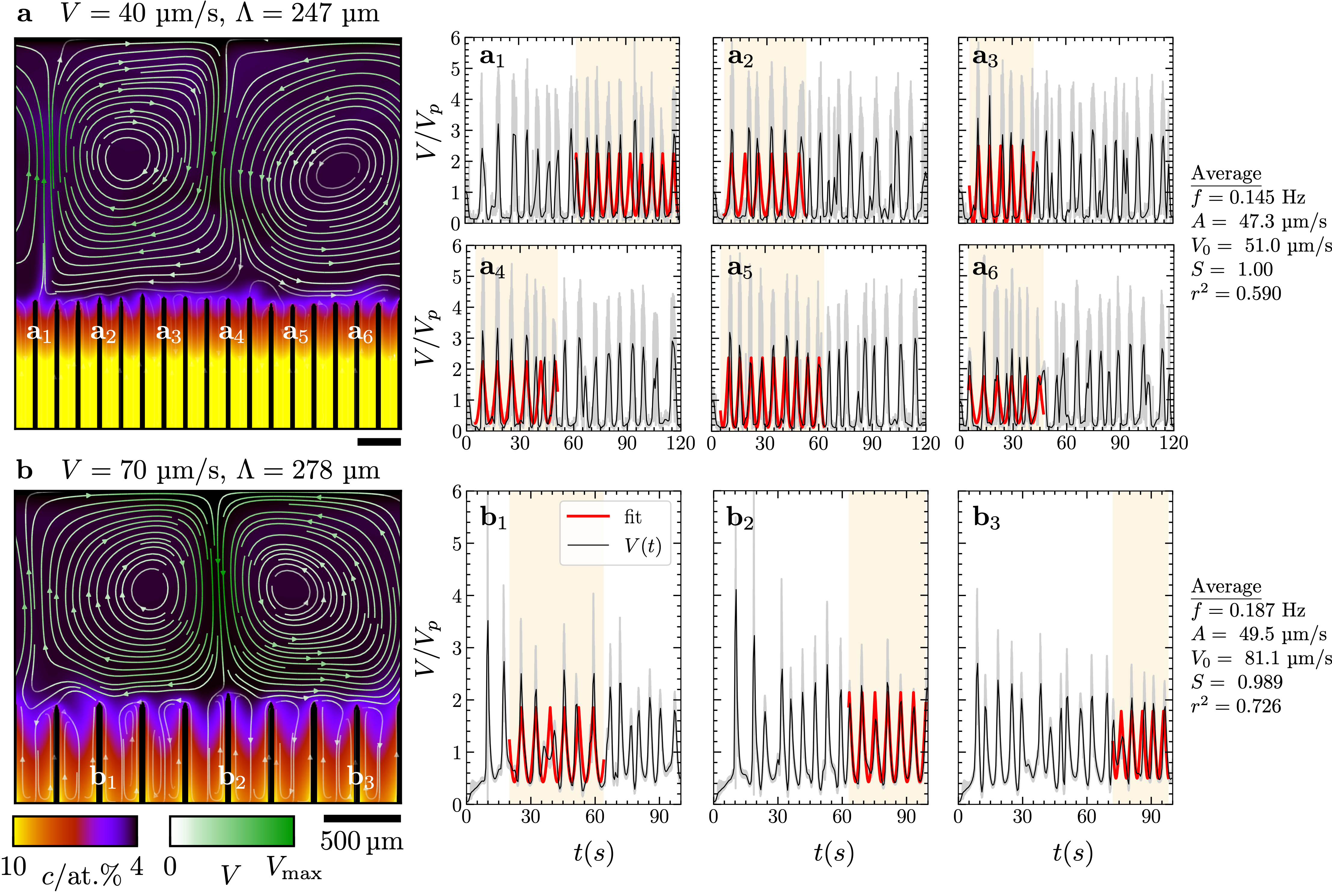}}%
\caption{
\underline{Noisy oscillations} for (a) $V=40~$\textmu m/s and $\Lambda=247~$\textmu m, and (b) $V=70~$\textmu m/s and $\Lambda=278~$\textmu m.
(left) Concentration field (color map) and fluid flow (streamlines, from opaque at $V/V_{\rm max}\geq0.2$ to 90\% transparent at $V=0$) at $t=$ (a) 120~s and (b) 100~s.
(right) Tip velocities of selected needles, showing raw (gray) and smoothed (black) $V(t)$ as well as fitted function (thick red) and range (shaded background).
  \label{fig:noisy}
}%
\end{figure}

\underline{Noisy oscillations (high $\Lambda$)}. 
As shown in Figure~\ref{fig:noisy}, at high primary spacings, or across all stable spacings for low $V_p$, the velocities deviate more prominently from the prototypical behaviors illustrated in Fig.~\ref{fig:functions}.
Figure~\ref{fig:noisy}a shows the worst encountered fit with $r^2=0.590$ and Fig.~\ref{fig:noisy}b a case closer to the edge of the transition between the ``sustained'' and ``noisy'' regions of Fig.~\ref{fig:map}.
While the $V(t)$ signals are clearly noisy and cannot be assimilated to any clear periodic function, both of them still exhibit an emerging frequency close to that of the sustained oscillations, namely with $0.134<~$f/Hz$~<0.255$ (i.e., periods between 3.92 and 7.46~s) across the investigated range of noisy oscillations.
This behavior remains quite different from the burst-like growth encountered for CMSX4 simulations at low $V_p$ \cite{isensee2022convective}.
All cases of noisy oscillations exhibit prominent convection rolls in the liquid phase. 

\section{Summary and Perspectives}

We investigated the conditions of occurrence and characteristics of buoyancy-driven oscillations of primary dendrite growth velocities in directional solidification using multiscale dendritic needle network (DNN) simulations.
While it was previously studied experimentally \cite{reinhart2020impact} and computationally \cite{isensee2022convective} for a Ni-based alloy, here we reproduce comparable behaviors using a binary Al-4at.\%Cu alloy.
These results suggest that this phenomenon may be general to any binary alloy, provided the appropriate $(G,V_p,\Lambda)$ conditions. 
We confirmed that the the ratio between average fluid velocity $\overline V$ and pulling velocity $V_p$ provides a reasonable indicator of the oscillatory regime (namely: damped, sustained, or noisy).
However, the threshold values separating regimes differ from those identified for CMSX4 under different processing conditions \cite{isensee2022convective}.
Here, the transition from damped to sustained oscillations occurs together with the onset of appearance of convection vortices in the fluid.
This is also in contrast with previous results for CMSX4 \cite{reinhart2020impact,isensee2022convective}, exhibiting well-established convection rolls in both regimes.
Our results also did not capture the further expected transition from sustained to damped oscillations when increasing $V_p$ (i.e., cooling rate \cite{reinhart2020impact,isensee2022convective}), perhaps because it occurs at higher $V_p$ than those explored here.
Moreover, the effect of the null velocity imposed at the solid-liquid interface -- in contrast, e.g., to imposing the growth velocity, accounting for the solid-liquid density change, or tracking the solid fraction into a mushy region -- remains to be investigated.

Some open questions remain on buoyancy-driven oscillations in directional solidification. 
Among other things, it remains unclear whether the phenomenon is relevant to bulk samples or is promoted by the 2D configuration (or quasi-2D thin-sample confinement in experiments).
Ongoing work, directly following up from the present study, will clarify the dependence of oscillatory behavior upon different alloy parameters (such as diffusion coefficient $D$ and partition coefficient $k$).
Scanning a broader range of alloy parameters and processing conditions will allow us to extract more meaningful scaling laws for the oscillation characteristics, and to establish general rules governing the stability of dendritic fronts in directional solidification processes.

\ack
This study was supported by the Spanish Ministry of Science and Innovation through the Mar\'ia de Maeztu seal of excellence of IMDEA Materials Institute (CEX2018-000800-M) and a Ram\'on y Cajal Fellowship (RYC2019-028233-I).

\section*{References}

\bibliographystyle{iopart-num} 
\bibliography{references}%

\end{document}